\documentclass[aps,showpacs,english,prl,twocolumn]{revtex4}
\usepackage{mathrsfs}
\pdfoutput=1
\usepackage{natbib}
\usepackage{graphicx}
\usepackage{amsmath, amssymb}
\usepackage{babel}

\begin{document}

\title{ Possibility of a zero-temperature metallic phase in granular two-band superconducting films }

\author{Bojun Yan}

\author{Tai-Kai Ng}

\affiliation{Department of Physcis, Hong Kong University of
Science and Technology, Hong Kong, People's Republic of China}

\date{today}

\date{Dec. 2011}


\begin{abstract}
 A variational approach is used to study the superconductor-insulator transition in two-band granular superconducting films using a resistance-shunted
 Josephson junction array model in this letter. We show that a zero-temperature metallic phase may exist between the superconducting and insulator
 phases which is absent in normal single band granular superconducting films. The metallic phase may be observable in some dirty pnictide superconductor
 films.

\end{abstract}

\maketitle

 Intensive studies had been devoted to the problem of superconductor-insulator (SI) transition in low-$T_c$ thin films. These systems undergo phase
 transitions from superconductor to insulator as a function of disorder, film thickness as well as external magnetic fields\cite{MPAF9023}. The SI
 transition is usually modelled by a Josephson junction array model, expressed in terms of the phases of the superconductor order parameter $\theta_i$'s
 on superconducting grain $i$'s. The Hamiltonian describing the system consists of the Josephson coupling between superconducting grains
 $\sim J\cos(\theta_i-\theta_j)$ where $(i.j)$ are nearest neighbor sites, and the charging energy $\sim {C\over2}(\dot\theta_i-\dot\theta_j)^{2}$. The
 system is in a superconducting phase if the Josephson term dominates, and is in the insulator phase if the charging energy dominates. It has been
 proposed by different authors that a dissipative term arising from coupling between superconducting grains and a dissipative metallic bath may also be important
 in describing the SI transition (shunted Josephson array model)\cite{CIKA8603,CIKZ8783,KMKC}. In particular, a zero-temperature metallic phase between superconductor and insulator phases may be stabilized by dissipation.

  The physical reason behind the metallic phase is as follows: Imagine first a state dominated by charging energy. In this case the metallic bath would
  screen the Coulomb potential, leading to a weakening of charging energy and drives the system towards a metallic phase if the resistance is small
  enough ($R<R_{cI}$)\cite{asos}.  Alternatively, the coupling of Cooper pairs in the superconducting phase to a dissipative environment suppresses
  coherent tunnelling of Cooper pairs between grains owing to the Calderia-Leggett effect\cite{caldeira-leggett} and superconducting coherence is destroyed if $R>R_{cS}$. As a result a metallic phase between the superconducting and insulating phases may exist if $R_{cS}<R<R_{cI}$. The metallic phase, if exist, is a new phase of matter because of participation of incoherent boson (Cooper pairs) in low temperature transports which is absent in usual metals. Experimentally the zero-temperature metallic phase in single band superconducting films has not been found to exist so far in the absence of external magnetic fields, consistent
  with a theoretical finding that $R_{cI}<R_{cS}$ in single-band superconductors\cite{NL0109}.

  More recently, superconductors with more than one order parameters, i.e., the multi-band superconductors\cite{SMW5952}
  have raised attention in the physics community. Examples of multi-band superconductors include MgB$_{2}$\cite{N0110}
  and the pnictide superconductors\cite{Kamihara}. It is interesting to see whether a metallic phase may exist more easily between the SI-transition in
  these materials. This is the purpose of this letter.

  Using a variational approach, we consider in this letter the superconductor-insulator transition in two-band $s$ (and $s_{\pm}$)-wave
  superconducting films where the possibility of an intermediate metallic phase is investigated. We show that contrary to
  the case of single-band superconductors, a physically realizable condition for the zero-temperature intermediate metallic
  phase is found for these systems. We propose that the metallic phase may be observable in some recently discovered disordered
  pnictide superconductors\cite{FW1101}.
  \begin{figure}[h]
  \begin{center}
  \includegraphics[width=86mm,height=30mm]{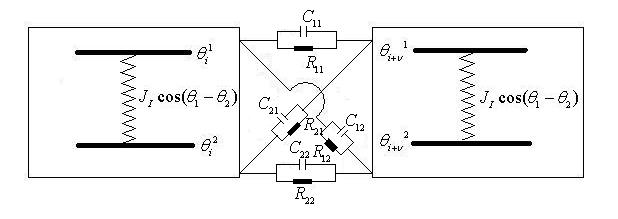}
  \end{center}
  \caption{A qualitative sketch of our two band Josephson junction model. The two bands on different grains are connected by
  a capacitor, a resistor and Josephson coupling (not shown in the sketch). The two bands on the same grain are connected
  by in-grain inter-band Josephson coupling $J_I$}
  \end{figure}

  We start with the phase-action which is a generalization of the phase action used to study superconductor-insulator transition in one-band systems\cite{NL0109, CIKA8603, CIKZ8783}. The system is schematically sketched in Fig.1. The action describes a resistance-shunted Josephson network of two-band superconductor grains and is given in imaginary time by $S=S_{\theta}+S_{\rm diss}$, where

 \begin{eqnarray}
 \label{scompact2} S_{\theta} &&= \sum_{i,\nu,a,b}\int^{\beta}_0 d\tau[{1\over 2}{C_{ab}} (\Delta_\nu\dot\theta_i^{ab})^{2}
 - J_{ab}\cos(\Delta_\nu\theta^{ab} _i)]\nonumber\\
 & & +J_I\sum_i\int^{\beta}_0 d\tau\cos(\theta^1_i-\theta^2_i)
\end{eqnarray}
 is the phase action without the dissipative term. $a,b=1,2$ represent the two different bands in a grain, and $\theta^{a}_i$
 is the phase of band $a$ superconducting order parameter in grain $i$. $\Delta_\nu\theta^{ab}_i=\theta^{a}_i-\theta^{b}_{i+\nu}$
 represents the phase difference between band $a$ and $b$ superconducting order parameters in neighboring grains $i$, $i+\nu$, respectively and
 $J_{ab}>0$ is the corresponding Josephson coupling energy. ${{1}\over{2}}C_{ab}(\Delta_\nu\dot\theta^{ab}_i)^2$ represents
 the charging energy arising from charge imbalance between band $a$ and band $b$ electrons on grain $i$ and $i+\nu$ respectively, where $C_{ab}$
 is the corresponding capacitance. $J_I$ is the in-grain inter-band Josephson coupling which favors $\theta^2_i=\theta^1_i+\pi$ for $J_I>0$,
 leading to a $s_{\pm}$ superconductor and favors $\theta^1_i=\theta^2_i$ for $J_I<0$ ($s$-wave superconductor).

 \begin{eqnarray}
 \label{sdiss}
 S_{\rm diss} && ={Q^2\over2}\sum_{i,\nu,a,b}\int^{\beta}_0 d\tau\int^{\beta}_0 d\tau' \nonumber\\
  && \times\alpha_{ab}^{\rm o}(\tau-\tau')\sin^{2}[{{D_{\nu}\theta^{ab}_i{(\tau)}-D_{\nu}\theta^{ab}_i{(\tau')}}\over 2Q}]
\end{eqnarray}
 where $\alpha_{ab}^{\rm o}(\tau)={(h/{4e^{2}R_{ab}^{\rm o}})}{[T/{\sin(\pi{T}\tau)}]^2}$. $R_{ab}^{\rm o}$ is the resistance between
 band $a$ and band $b$ electrons on grains $i$ and $i+\nu$, respectively (see Fig.(1)) and $Q=2$ is the charge of a Cooper pair.
 $D_\nu\theta_i^{12}=D_\nu\theta_i^{21}=[\Delta_\nu\theta_i^{11}+\Delta_\nu\theta_i^{22}]$,
 $D_\nu\theta_i^{aa}=[{{3}\over{2}}\Delta_\nu\theta_i^{aa}+{{1}\over{2}}\Delta_\nu\theta_i^{\bar a\bar a}]$, where
 $\bar1({\bar2})=2(1)$.
 $S_{\rm diss}$ is derived phenomenologically from a multi-band resistance network model represented by Fig.1. The details
 of the derivation can be found in the supplementary materials.

 To simplify calculation we shall consider the grains forming a two-dimensional square lattice with $J_{12}=0$ in our
 following analysis. With the later condition the $s$ and $s_{\pm}$ superconductors can be transformed to each other by
 simply shifting $\theta^2_i\rightarrow\theta^2_i+\pi$. The main effect of $J_{12}$ is to renormalize
 $J_I\rightarrow J_I-zJ_{12}$ where $z$ is the lattice co-ordination number and is not going to affect our
 conclusion in renormalization-group sense.

 Due to the compactness of the phase field ($e^{i(\theta+2n\pi)}=e^{i\theta}$), the phase variables $\theta^a_i(\tau)$ can be
 decomposed into a periodic part and a winding number contribution,
\[
\theta^{a}_i(\tau)={{2\pi{n^{a}_i}\tau}\over{\beta}}+\theta^{a}_{i0}(\tau)
\]
 where $\theta^{a}_{i0}(\beta)=\theta^{a}_{i0}(0)$ and $n^{a}_i$ can be any arbitrary
 integer (winding number). With this decomposition the phase action becomes
 \begin{eqnarray}
 \label{sbosonsep}
 S_{\theta} & \rightarrow & {{2\pi^2}\over{\beta}}\sum_{i,\nu,a,b}C_{ab}{{\Delta_\nu n^{ab}_i}^2}+
 \sum_{i,\nu,a,b}{C_{ab}\over2}\int^{\beta}_0d\tau(\Delta_\nu\dot\theta^{ab}_{i0})^2 \nonumber\\
 && -\sum_{i,\nu,a}J_{aa}\int^{\beta}_0 d\tau\cos[\Delta_{\nu}\theta^{aa}_{i0}+{2\pi\tau\over\beta}\Delta_{\nu}n^{aa}_{i}]  \nonumber\\
 && +J_I\sum_i\int^{\beta}_0d\tau\sum_{\nu}\cos[\Delta\theta_{i0}+{{2\pi\tau}\over\beta}\Delta n_{i}].
\end{eqnarray}
 where $\Delta_{\nu}n^{ab}_{i}=n^a_i-n^b_{i+\nu}$, $\Delta\theta_{i0}=\theta^1_{i0}(\tau)-\theta^2_{i0}(\tau)$ and
 $\Delta n_i=n^1_i-n^2_i$ and
\begin{eqnarray}
\label{sdisssep}
 S_{\rm diss} & \rightarrow & \sum_{i,\nu,a,b}{Q\pi\over4R_{ab}}|D_{\nu}n^{ab}_i|+
 {1\over 8}\sum_{i,\nu,a,b}\int^{\beta}_0 d\tau\int^{\beta}_0 d\tau'
   \nonumber\\
  & & \alpha_{ab}^{\rm o}(\tau-\tau')\cos[{2\pi(\tau-\tau')\over Q\beta}D_{\nu} n^{ab}_i]  \nonumber\\
 & & \times[D_{\nu}\theta^{ab}_{i0}(\tau)-D_{\nu}\theta^{ab}_{i0}(\tau')]^2.
 \end{eqnarray}
 where $D_\nu n_i^{ab}$ is defined in the same way as $D_\nu\theta_i^{ab}$ with $\Delta_{\nu}\theta_i^{ab}\rightarrow\Delta_{\nu}n_i^{ab}
 =n^a_i-n^b_{i+\nu}$. We have assumed strong dissipation and keep only to second order terms of  $\Delta_\nu\theta^{aa}_{0i}$ in $S_{\rm diss}$ for simplicity\cite{NL0109}.

  To proceed further we employ a variational approach\cite{NL0109}. We consider a trial action $S_{\rm trial}=S^P_{\rm trial}+S^n_{\rm trial}$
  where the periodic and the winding number contributions to $\theta$ are
  decoupled.

  \begin{eqnarray}
 \label{trialtheta}
  S^P_{\rm trial} &=& \sum_{i,\nu,a,b}\int^{\beta}_{0}d\tau[{C_{ab}\over 2}(\Delta_\nu\Dot{\theta}^{ab}_{i0})^2+{J_{aa}^{\rm eff}\over 2}(\Delta_\nu\theta^{aa}_{i0})^2]  \nonumber\\
  & & +{J_I^{\rm eff}\over 2}\sum_{i}\int^{\beta}_0 d\tau{(\Delta\theta_{i0})^2}+  \nonumber\\
  & & {1\over 8}\sum_{i,\nu,a,b}\int^{\beta}_0d\tau\int^{\beta}_0d\tau'\alpha^{\rm{eff},o}_{ab}(\tau-\tau')\times \nonumber\\
  & & {[D_\nu\theta^{ab}_{i0}(\tau)-D_\nu\theta^{ab}_{i0}(\tau')]^2}
 \end{eqnarray}
 is an effective action describing Gaussian fluctuations of the periodic phases around the saddle point $\theta^1_{i0}(\tau)=0$,
 $\theta^2_{i0}(\tau)=0(\pi)$ for $J_I<(>)0$ and
 \begin{eqnarray}
 \label{trialwinding}
 S^n_{\rm trial}& = & {2\pi^2\over\beta}\sum_{i,\nu,a,b}C_{ab}{\Delta_\nu n^{ab}_i}^2-\beta J_I^{MS}\sum_i\delta(n_i^1-n_i^2)  \nonumber\\
 & + & \sum_{i,\nu,a,b}\left[{Q\pi\over4R^{\rm{eff},o}_{ab}} |D_\nu n^{ab}_i|-\beta
 J_{aa}^{MS}\delta(\Delta_{\nu}n^{aa}_i)\right]\nonumber\\
 \end{eqnarray}
 in an effective action for the winding number field. $S^n_{\rm trial}$ is a generalized absolute solid-on-solid model
 (ASOS) for two species of winding numbers with additional $\beta J^{MS}$ terms originating from superconductivity. $J_{aa}^{\rm eff}$,$J_I^{\rm eff}$, $J_{aa}^{MS}$, $J_I^{MS}$,
 $\alpha^{\rm eff,o}_a(\tau)/\alpha(\tau)=R_a^{\rm o} /R^{\rm eff,o}_a$ are variational parameters to be determined by minimizing the free energy of the system given
 approximately by $F=F_0+\langle S-S_{\rm trial}\rangle$, where $F_0$ is the free energy computed using $S_{\rm trial}$ and $\langle...\rangle$ denotes
 averages taken with respect to $S_{\rm trial}$.

 The different phases can be identified in our trial action as follows: First we note that $S^P_{\rm trial}$ describes a stable superconducting
 phase as long as the phase stiffness's satisfy $J^{\rm eff}_{aa}>0$. The nature of the $J^{\rm eff}_{aa}=0$ (non-superconducting) state is determined by $S^n_{\rm trial}$ which describes two different possibilities. For small $R_{ab}^{\rm eff,o}$'s, the system is in a ``smooth" phase where fluctuations in $n^a_i$'s are
 suppressed and charges become mobile. The system is in a metallic phase. For large $R_{ab}^{\rm eff,o}$'s
 $n^a_i$'s at different sites fluctuate violently (rough phase) and charge fluctuations are suppressed. The
 system is in the insulator phase\cite{asos}.

 Minimizing the free energy we obtain after some lengthy algebra the mean-field equations
 \begin{eqnarray}
 \label{mf}
 R^{\rm eff,o}_{ab} & = & R_{ab}^o  \\ \nonumber
 J^{MS}_{aa} & = & J_{aa}e^{-\langle|\Delta_{\nu}\theta_{i}^{aa}|^2\rangle}  \\
 \nonumber
 J^{MS}_I & = & |J_I|e^{-\langle|\Delta\theta_{i}|^2\rangle}  \\ \nonumber
 J^{\rm eff}_{aa} & = & J^{MS}_{aa}P_{n}^{aa}(0)  \\ \nonumber
 J^{\rm eff}_I & = & J^{MS}_IP_{n}^I(0),
 \end{eqnarray}
 where $P_{n}^{aa}(m)=\langle\delta(m-|\Delta_{\nu}n_i^{aa}|)\rangle_{\rm S^n}$
 and $P_{n}^I(m)=\langle\delta(m-|n_i^1-n_i^2|)\rangle_{\rm S^n}$
 are the probabilities that the integer differences $|\Delta_{\nu}n_i^{ab}|=m$ and
 $|n_i^1-n_i^2|=m$ in $S^n_{\rm trial}$, respectively.
 \begin{subequations}
 \label{master}
 \begin{equation}
 \label{master1}
 \langle|\Delta_{\nu}\theta_{i}^{aa}|^2\rangle={1\over \beta N d}\sum_{i\omega_n ,\vec{k}}
 {\gamma(\vec{k})\over2}{a_{\bar{a}\bar{a}}\over a_{11}a_{22}-a^2_I},
 \end{equation}
 where $\bar{1}(\bar{2})=2(1)$ and
 \begin{equation}
 \label{master4}
 \langle|\Delta\theta_i|^2\rangle={1\over 2\beta N d}\sum_{i\omega_n ,\vec{k}}
 {a_{11}+a_{22}+2a_I\over a_{11}a_{22}-a^2_I},
 \end{equation}
 \end{subequations}
 where
 \begin{subequations}
 \label{alpha}

 \begin{equation}
 \label{a1}
 a_{bb}=\left({J^{\rm eff}_{bb}}+{{1}\over{2}}(\alpha_{bb}+{\alpha_{12}})|\omega_n|\right)\gamma(\vec{k})+J^{\rm eff}_I
 \end{equation}
 $(b=1,2)$ and
 \begin{equation}
  \label{a3}
  a_I=-J^{\rm eff}_I+{\alpha_{12}\over2}|\omega_n|\gamma(\vec{k})
 \end{equation}
 \end{subequations}
 where $\alpha_{ab}=h/(4\pi e^2R_{ab})$. The resistance $R_{ab}$'s are given by $R_{aa}^{-1}={{1}\over{2}}(3R^{o-1}_{aa}-R_{\bar a\bar a}^{o-1})$, and
 $R_{12}^{-1}=R_{12}^{o-1}+{{3}\over{4}}(R_{11}^{o-1}+R_{22}^{o-1})$ where $\alpha_{ab}^o=h/(4\pi e^2R_{ab}^o)$. This
 rather complicated form of resistance is a result of appearance of $D_\nu\theta$ terms in $S_{\rm diss}$.
  $\gamma (\vec k)=4(\sin^2(k_x/2)+\sin^2(k_y/2))$ is the geometric factor of 2D square
 lattice.

 The phase diagram of the system is determined by solving the above equations numerically. Notice that the superconducting transition given by
 $J^{\rm eff}_{aa}=0$ is determined by $S^P_{\rm trial}$ only and is independent of $S^n_{\rm trial}$ as long as $P_{n}^{aa}(0)$ and $P_{n}^I(0)$ are
 nonzero. Similarly, the metal to insulator transition is determined by $S^n_{\rm trial}$ only (rough or smooth phase) when $J^{\rm eff}_{a}=0$. In
 Fig.2 we present the resulting phase diagram for the symmetric case $J_{11}=J_{22}$, $C_{11}=C_{22}$ and $R_{11}=R_{22}$ for two different values of
 $J_I$. We note that a metallic phase is found in a narrow region of parameter space when $J_I$ is small enough, contrary to the single-band case where no
 metallic phase is found.

 \begin{figure}[h]
\begin{center}
\includegraphics[width=86mm,height=65mm]{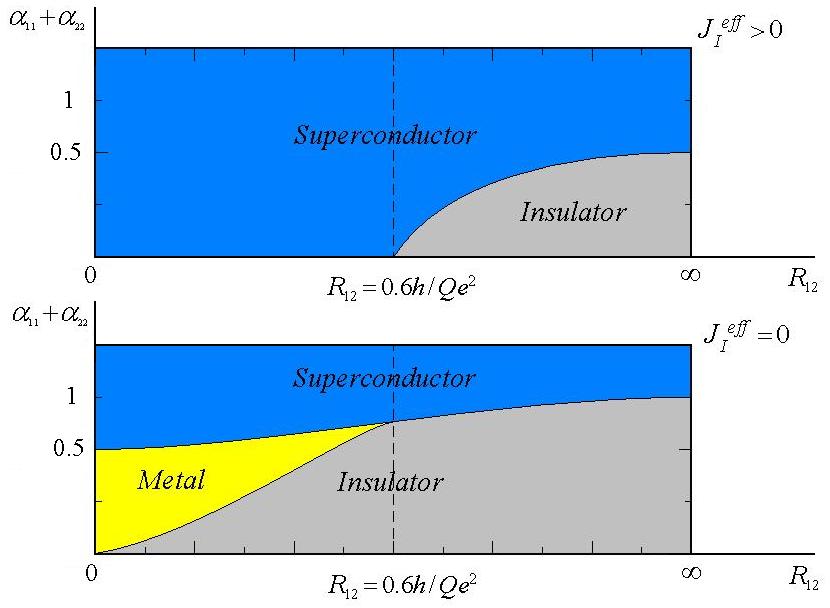}
\end{center}
 \caption{Phase diagram of the two band system for different values of $\alpha_{11}+\alpha_{22}$ versus $R_{12}$ for two values of
 $J_I$ with $J^{\rm eff}_I\neq0$ in the upper panel and $J^{\rm eff}_I=0$ in the lower panel. A metallic phase is found in a narrow region of parameter
 space when $J^{\rm eff}_I=0$ and $R_{12}<0.6h/Qe^2$}
\end{figure}

  To understand the phase diagram, we observe first that the phase diagram is divided into two regimes, (i)$J^{\rm eff}_I\neq0$ (upper panel) and
  (ii)$J^{\rm eff}_I=0$ (lower panel). The effective Josephson coupling between the two bands is nonzero in the first regime and it is easy to show
  from Eqs.\ (\ref{mf}-\ref{alpha}) that the two-band superconductor
  becomes effectively like a single band superconductor at low energy $\omega<<J^{\rm eff}_I$ in $S^P_{\rm trial}$. Correspondingly, $S^n_{\rm trial}$  becomes an effective one band model with $n_i^1\equiv n_i^2=n_i$ at temperature $T\rightarrow0$ because of the
  $\beta J_I^{MS}\sum_i\delta(n_i^1-n_i^2)$ term, i.e.
   \[
  S^n_{\rm trial}\rightarrow{2\pi^2C_{t}\over\beta}\sum_{i}{\Delta_\nu n_i^2}+{Q\pi\over4R_{\rm eff}}\sum_{i,\nu}|\Delta_\nu n_i|
  \]
  when superconductivity is destroyed, where $C_{t}=C_{11}+C_{22}$ and $R_{\rm eff}^{-1}=R^{-1}_{11}+R^{-1}_{22}+2R^{-1}_{12}$. The system is at a
 roughening (insulator) phase when $R_{\rm eff}>R_{cI}=0.6(h/Qe^2)$\cite{asos} and a metallic phase exist only if $J_{aa}^{\rm eff}=0$ at a finite region of
 resistances $R_{cI}>R_{\rm eff}>R_{cS}$. For single-band superconductors, $R_{cS}=(h/Qe^2)>R_{cI}$\cite{NL0109}, and an intermediate metallic phase
 cannot exist in this case.

   We next consider regime (ii) where $J^{\rm eff}_I=0$. First it is straightforward to show that $J^{\rm eff}_I\neq0$ as long as $J^{\rm eff}_{aa}\neq0$,
   indicating that the system behaves always like an effective one-band system at low enough energy in the superconducting state. The situation
   is different if superconductivity is destroyed. Substituting $J^{\rm eff}_{aa}=0$ into Eq.\ (\ref{master4}), we obtain a self-consistent equation for $J^{\rm eff}_I$,
   \begin{equation}
   \label{ji}
    J^{\rm eff}_I=|J_I|P_{n}^I(0)\exp\left(-{1\over 2\beta N d}\sum_{i\omega_n ,\vec{k}}{1\over J^{\rm eff}_I+\tilde{\alpha}_I|\omega_n|\gamma(\vec{k})}\right),
    \end{equation}
   where $\tilde{\alpha}_I={\alpha_{11}\alpha_{22}+\alpha_{12}(\alpha_{11}+\alpha_{22})\over \alpha_{11}+\alpha_{22}+4\alpha_{12}}$.

   Equation\ (\ref{ji}) is solved numerically where we find that the equation has a non-zero solution only when $|J_I|>J^c_I(\tilde{\alpha}_I,J_{11})$,
   which is a number depending on $\tilde{\alpha}_I$ and roughly proportional to $\max(J_{11},J_{22})$, the transition from the $J^{\rm eff}_I\neq0$ to
   $J^{\rm eff}_I=0$ state is a first order phase transition. The phase diagram determined by\ (\ref{ji}) is
   provided in the supplementary material.


    This interesting result suggests that although the superconducting state behaves always like an effective one-band superconductor at low enough energy, there
    exists two kinds of non-superconducting states. The non-superconducting state is effectively one-band like when $J^{\rm eff}_I\neq0$ and two-band
    like when $J^{\rm eff}_I=0$. We find that an intermediate metallic phase may exist in the two-band like non-superconducting state.

  To see how this can occur we consider Eq.\ (\ref{master1}) with $J^{\rm eff}_I=0$. In this case we obtain
  \begin{equation}
  \label{effone}
   J^{\rm eff}_{aa}\sim J_{aa}P_{n}^{aa}(0)\exp\left(-{1\over 2\beta N d}\sum_{i\omega_n ,\vec{k}}
 {1\over J^{\rm eff}_{aa}+\tilde{\alpha}_{aa}|\omega_n|}\right),
  \end{equation}
  where
  \[
  \tilde{\alpha}_{aa}={h\over4\pi e^2}\left({1\over R_{aa}}+{1\over
  R_{\bar{a}\bar{a}}+2R_{12}}\right), \]
   corresponding to a single-band superconductor with effective resistance $R_{\rm eff}^{-1}=R_{aa}^{-1}+(R_{\bar{a}\bar{a}}+2R_{12})^{-1}$, which is
   the effective resistance obtained from the resistance network model shown in Fig.(1). The SI transition is determined by Eq.\ (\ref{effone}) and
   $S^n_{\rm trial}$. To see the plausible existence of metallic phase, we examine the limit $R_{12}^{\rm o}\sim R_{12}\rightarrow0$. In this limit, a
   long-range order of $n^{1,2}_i$'s are built up in the winding number action $S^n_{\rm trial}$ because $n_i^1\equiv n_{i+\nu}^2 \forall i$, and the
   system is always in the smooth phase. A metallic phase exists as long as $R_{\rm eff}\rightarrow R_{11}R_{22}/(R_{11}+R_{22})>h/Qe^2$ where
   $J^{\rm eff}_{aa}\rightarrow0$.

    The window for the existence of metallic phase narrowed down when $R_{12}$ increases as shown in Fig.(2) lower panel.
    Notice that the winding number field is basically controlled by $R_{12}$ when $R_{11}$ and $R_{22}$ are large, so for a metallic phase to occur, we
    generally require $R_{12}$ to be smaller than $0.6h/Qe^2$. For small $R_{12}$, the superconductor-insulator transition is governed by
    $\alpha_{11}+\alpha_{22}$. For the superconducting stiffness to vanish, we require $\alpha_{11}+\alpha_{22}\leq0.5$.

    The metallic phase, if exists, is a new state of matter with incoherent bosons participating in low temperature transports. The state is described
    by a Ginsburg-Landau (GL) theory with vanishing phase-stiffness\cite{ngepl}. A preliminary analysis of the GL theory
    indicates that the system is a diamagnetic metal with unusual low-temperature magneto-transport behaviors\cite{ngepl}.


 To conclude, we re-examine the problem of SI transition in this paper for two-band superconductors, and raise again the question of plausible existence
 of metallic phase. Within a resistance-shunted Josephson network array model, we show that intermediate metallic phase between superconductor-insulator
 transition may exist for two-band superconducting films if the inter-band Josephson coupling $J_I$ and inter-band dissipative resistance term $R_{12}$
 are small enough. Physically, the more complicated circuit network structure for two-band superconductors (Fig.1) gives rise to the possibility that the
 effective dissipation responsible for screening and quantum dissipation are coming from different resistance channels which is not possible for single-band
 superconductors. With the recent advancements of research in Iron pnictide and other multi-band superconductors, we believe that this new metallic phase of
 matter may be reachable in the near future\cite{FW1101}.

   This work is supported by HKRGC through grant HKUST3/CRF09.


\newpage

\section{Supplementary materials}

 Here we show how the dissipative term \ (\ref{sdiss}) can be derived from a straightforward generalization of the dissipative term for one-band system
 to two-band system. We assume phenomenologically that a metallic component exists in the system and the
 dissipative term can be derived from a Hamiltonian with
 tunnelling and capacitance energy between grains,
\begin{equation}
H=\sum_{m,a}H_{m}^a+\sum_{a,b}\left(H_T^{ab}+H_Q^{ab}\right)
\end{equation}
 where $m=L,R$ and $a,b=1,2$ are the grain and band indices, respectively.
\begin{subequations}
\begin{equation}
\label{411a}
H^{a}_{m}=\sum_{\sigma}\int dx_{m}\hat\Psi_\sigma^{\dagger a}(x_m)[\epsilon^a_{m}(-i\nabla)]\hat\Psi_\sigma^a(x_{m})
\end{equation}
describes non-interacting electrons in grain $m$, band $a$ where $\sigma$ is the spin index, and
\begin{equation}
\label{411b}
H_T^{ab}=\sum_{\sigma}\int d x_L d x_R T^{ab} (x_L,x_R)\hat\Psi_\sigma^{\dagger a}(x_L) \hat\Psi_\sigma^{b}(x_R)+h.c.
\end{equation}
describes tunneling of electrons between grain $L$, band $a$ and grain $R$, band $b$ and
\begin{equation}
\label{411c}
H_Q^{ab}={1\over{8C}}(Q_R^a-Q_L^b)^2
\end{equation}
is the charging energy associated with charge imbalance between the grains where
\end{subequations}
\begin{equation}
\label{412}
Q_{m}^{a}=e\sum_{\sigma}\int d x_{m}\hat\Psi_\sigma^{\dagger a}(x_{m}) \hat\Psi_\sigma^{a}(x_{m})
\end{equation}
is the total electric charge in grain $m$, band $a$. The corresponding action at imaginary time is
\begin{eqnarray}
\label{413}
S&&=\sum_{a,\sigma}\int_0^{\beta}d\tau\{{\int d x_L{\bar\Psi_\sigma^a(x_L)\partial\tau\Psi_\sigma^a(x_L)}}\nonumber\\
&&+\int d x_R{{\bar\Psi_\sigma^a(x_R)\partial\tau\Psi_\sigma^a(x_R)}}\}+H
\end{eqnarray}
To derive $S_{\rm diss}$ we  first apply a Stratonovich-Hubbard
transformation on $H_Q$ to obtain
\begin{eqnarray}
\label{414}
S&&\rightarrow\sum_{m,a,\sigma}\int_0^\beta\int d x_{m} \bar\Psi_\sigma^a(x_m)\{\partial\tau+\epsilon^a_{m}(-i\nabla)\nonumber\\
&&+(-1)^{s_m}({{ie}\over{2}})[V_{aa}\nonumber\\
&&+({{1+(-1)^{s_m}}\over{2}})V_{a\bar a}+({{1-(-1)^{s_m}}\over{2}})V_{\bar a a}]\}\Psi^a_\sigma(x_m)\nonumber\\
&&+\sum_{a,b,\sigma}\int_0^\beta d\tau\int dx_{L}dx_{R}\nonumber\\
&&\times[T^{ab}(x_L,x_R)\bar\Psi^a_\sigma(x_L)\Psi^b_\sigma(x_R)+c.c] \nonumber\\
&&-\sum_{a,b}\int_0^\beta d\tau[{{CV^2_{ab}}\over{2}}]
\end{eqnarray}
where $m=L,R$, $s_L=0$ and $s_R=1$ and $\bar{1}(\bar{2})=2(1)$.

 Writing $V_{ab}=\dot{\theta}^a_R-\dot{\theta}^b_L$, where $\bar{L}(\bar{R})=R(L)$, the electric potential $V_{ab}$'s can be
 absorbed by a gauge transformation
\begin{equation}
\label{415a}
 \Psi^a_\sigma(x_m,\tau)=e^{-i\theta_{m}^a+{{i}\over{2}}(\theta_{\bar m}^1+\theta_{\bar m}^2)}\tilde\Psi^a_\sigma(x_m,\tau).
\end{equation}
 where the tunnelling term becomes
\begin{eqnarray}
\label{416}
S_T&&\rightarrow\sum_{a,b,\sigma}\int_0^\beta d\tau\int dx_{L}dx_{R} \{T^{ab}(x_L,x_R)\nonumber\\
&&\times e^{i[{{1}\over{2}}(\Delta_\nu\theta^{ab}+\Delta_\nu\theta^{ba})+{{1}\over{2}}(\Delta_\nu\theta^{12}+\Delta_\nu\theta^{21})]}\nonumber\\
&&\times\bar\Psi^a_\sigma(x_L)\Psi^b_\sigma(x_R)+c.c\}
\end{eqnarray}
 where $\Delta_\nu\theta^{ab}=\theta_{L}^a-\theta_{R}^b$ $(a,b=1,2)$. To proceed further, we integrate out the
 fermionic fields and expand the tunnelling term to second order to obtain
\begin{equation}
\label{5117}
 S_{\rm eff}(\theta)=\sum_{a,b}\int_0^\beta d\tau[{C_{ab}\over2}\Delta_\nu\dot\theta^{ab}(\tau)^2+S_{\rm diss}^{ab}]
\end{equation}
 where
\begin{eqnarray}
\label{5118}
&&S_{\rm diss}^{ab}={1\over2}\sum_{\sigma}|T^{ab}|^2\int_0^\beta d\tau_1\int_0^\beta d\tau_2\nonumber\\
&&\{G_{L\sigma}^{a}(\tau_1-\tau_2)G_{R\sigma}^{b}(\tau_2-\tau_1)e^{i(D_\nu\theta^{ab}(\tau_2)-D_\nu\theta^{ab}(\tau_1))}\nonumber\\
&&+G_{R\sigma}^{b}(\tau_1-\tau_2)G_{L\sigma}^{a}(\tau_2-\tau_1)e^{i(D_\nu\theta^{ba}(\tau_1)-D_\nu\theta^{ba}(\tau_2))}\}\nonumber\\
\end{eqnarray}
 where $D_\nu\theta_i^{12}=D_\nu\theta_i^{21}=[\Delta_\nu\theta_i^{11}+\Delta_\nu\theta_i^{22}]$
 and $D_\nu\theta_i^{aa}=[{{3}\over{2}}\Delta_\nu\theta_i^{aa}+{{1}\over{2}}\Delta_\nu\theta_i^{\bar{a}\bar{a}}]$.

\begin{equation}
 \label{5119} G_{m\sigma}^a(\tau)={1\over{\beta V}}\sum_{i\omega,\vec k}{{e^{-i\omega_n
 \tau}}\over{i\omega_n-\epsilon_{m\sigma}^a(\vec k)}}=-{D_{m\sigma}^a(E_F)\pi T\over{\sin(\pi T\tau)}}
 \end{equation}
 is the free electron Green's function at imaginary time. $m=L,R$ and $D_{m\sigma}^a(E_F)$ is the density of states on the Fermi surface.
 Defining
 \begin{eqnarray}
 \label{5122} \alpha_{ab}^o(\tau) & = & \sum_{\sigma}|T^{ab}|^2D_{L\sigma}^a(E_F)D_{R\sigma}^b(E_F)({{\pi
 T}\over{\sin(\pi T\tau)}})^2    \nonumber\\
 & = & \left({h\over 4e^2R_{ab}^o}\right)\left({T\over\sin(\pi T\tau)}\right)^2
\end{eqnarray}
and put it back into \ (\ref{5118}) we obtain
\begin{eqnarray}
\label{417}
S_{\rm diss}&&\sim\sum_{a,b}\int_0^\beta d\tau_1\int_0^\beta d\tau_2\{\alpha_{ab}^o(\tau_1-\tau_2)\nonumber\\
&&\times\sin^2[{{D_\nu\theta^{ab}(\tau_1)-D_\nu\theta^{ab}(\tau_2)}\over{2Q}}]\}\nonumber\\
\end{eqnarray}
 which is the dissipation term we use in the main text.

  We attach here also the phase diagram determined by Eq.\ (\ref{ji}) with $J_{11}=J_{22}$. The line separating the $J_I^{\rm eff}=(\neq)0$ phases
 is a line of first order phase transition. we see that $J_{11}/J_I\sim\tilde{\alpha}_I={\alpha_{11}\alpha_{22}+\alpha_{12}(\alpha_{11}+\alpha_{22})\over
 \alpha_{11}+\alpha_{22}+4\alpha_{12}}$ at the transition.

 \begin{figure}[h]
\begin{center}
\includegraphics[width=85mm,height=60mm]{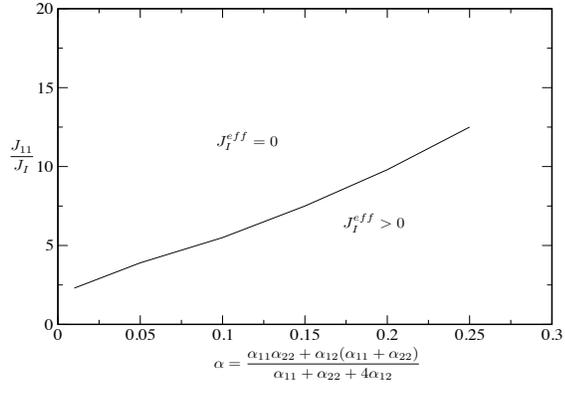}
\end{center}
\caption{Phase diagram for $J^{eff}_I$ for different values of
$J_I^{-1}$ versus $\tilde{\alpha}$. A first order phase transition
separates the $J_I^{eff}=0$ and $J_I^{eff}\neq0$ phases}
\end{figure}

\end{document}